\documentclass[aps,prd,preprint,nofootinbib,11pt]{revtex4}
\usepackage{amsfonts}
\usepackage{mathrsfs}
\usepackage{graphicx}
\usepackage{amsmath}
\usepackage{amssymb}
\usepackage{hyperref}
\usepackage{color}
\usepackage{subfigure}
\usepackage{epsfig}
\usepackage{graphicx}

\newcommand{\bqa}{\begin{eqnarray}}
\newcommand{\eqa}{\end{eqnarray}}
\newcommand{\beq}{\begin{equation}}
\newcommand{\eeq}{\end{equation}}
\newcommand{\ket}{\,\rangle}
\newcommand{\bra}{\langle \,}
\newcommand{\mL}{\mathcal{L}}

\begin{document}

\title{Determination of the up/down-quark mass within QCD sum rules in the scalar channel \\[0.7cm]}

\author{Fang-Hui Yin$^{1}$}
\author{Wen-Ya Tian$^{1}$}
\author{Liang Tang$^{1}$}
\email{tangl@hebtu.edu.cn}
\author{Zhi-Hui Guo$^{2}$}
\email{zhguo@seu.edu.cn}
\affiliation{$^1$ College of Physics and Hebei Key Laboratory of Photophysics Research and Application, Hebei Normal University, Shijiazhuang 050024, China\\
$^2$ School of Physics, Southeast University, Nanjing 211189, China}

\begin{abstract}
\vspace{0.3cm}
In this work, we determine up/down-quark mass $m_{q=u/d}$ in the isoscalar scalar channel from both the Shifman-Vainshtein-Zakharov (SVZ) and the Monte-Carlo-based QCD sum rules. The relevant spectral function, including the contributions from the $f_0(500)$, $f_0(980)$ and $f_0(1370)$ resonances, is determined from a sophisticated $U(3)$ chiral study. Via the traditional SVZ QCD sum rules, we give the prediction to the average light-quark mass $m_q(2 \, \text{GeV})=\frac{1}{2}(m_u(2 \, \text{GeV}) + m_d(2 \, \text{GeV}))=(3.46^{+0.16}_{-0.22} \pm 0.33) \, \text{MeV}$. Meanwhile, by considering the uncertainties of the input QCD parameters and the spectral functions of the isoscalar scalar channel, we obtain $m_q (2\, \text{GeV}) = (3.44 \pm 0.14 \pm 0.32) \, \text{MeV}$ from the Monte-Carlo-based QCD sum rules. Both results are perfectly consistent with each other, and nicely agree with the Particle Data Group value within the uncertainties.
\end{abstract}

\maketitle

\newpage

\section{Introduction}

The up- and down- quark masses  are fundamental parameters in Quantum Chromodynamics (QCD), and they play important roles in the phenomenological study of the Standard Model. However, they are not physical observables, due to the confinement nature of QCD. Usually one needs to adopt some QCD-based non-perturbative methods, such as lattice QCD, QCD sum rules, effective field theories, etc., to determine the quark masses. A recent review of the determinations of the light quark masses from lattice QCD can be found in Ref.\cite{Aoki:2019cca}.

The QCD sum rules, developed more than forty years ago by Shifman, Vainshtein and Zakharov (SVZ) \cite{Shifman, P.Col, Latorre:1985uy}, has some peculiar advantages in the study of hadron phenomenology. Its starting point in evaluating the properties of the concerned hadron is to construct the current, which possesses the important information about the concerned hadron, like quantum numbers, constituent quarks and gluons. With the currents one can then construct the two-point correlation function, which matches the QCD and phenomenological sides of sum rules. Within the QCD sum rules, in recent years there are remarkable progresses on the determination of the light quark masses\cite{Becchi:1980vz, Bijnens:1994ci, Jamin:2001zr, Steele:2001ga, Gupta:2003fn, Bodenstein:2013paa, Narison:2014vka, Dominguez:2018azt,Cherry:2001cj,Narison:2002hk,Maltman:2002sb}. Many of these works rely on the pseudoscalar condensates or the divergences of the axial-vector currents to determine the light quark masses from QCD sum rules. The golden cases to determine the $m_d+m_u$ and $m_d-m_u$ would be the divergence of the isovector axial-vector current  and the divergence of isovector vector current, respectively~\cite{Dominguez:1986aa,Dominguez:2018azt}. Alternatively the isoscalar scalar channel can be also used to calculate the $m_d+m_u$~\cite{Reinders:1981ww,Cherry:2001cj,Yuan:2017foa}. In this situation, on the phenomenological side one needs the isoscalar scalar spectral functions, which can contain the contributions from the scalar resonances $f_0(500)$($\sigma$), $f_0(980)$, $f_0(1370)$ and other heavier ones. Indeed there are serve debates on the roles of the different resonances played in the scalar QCD sum rules~\cite{Cherry:2001cj,Shakin:2001hc,Elias:1998bq,Shi:1999hm}, specially on the low lying states $f_0(500)$ and $f_0(980)$. In the present work, we attempt to address this issue by adopting the phenomenological spectral functions calculated within the unitarized $U(3)$ resonance chiral theory~\cite{Guo:2012ym,Guo:2012yt,Guo:2011pa}. The spectral functions are constructed by the unitarized scalar form factors, which are pure predictions once the scattering amplitudes are determined through the fits to scattering observables. In the complex energy plane,  the isoscalar scalar resonance poles for the $f_0(500)$, $f_0(980)$, $f_0(1370)$ are found, whose contributions are coherently included in the spectrum functions, together with two-meson continuum effects from the chiral effective field theory. The  prescription of the unitarized chiral spectrum functions  improves the Breit-Wigner description of the broad scalar $\sigma$ resonance~\cite{Yuan:2017foa}. Furthermore, there is no free parameter in our spectrum functions, since they can be predicted from the scattering observables~\cite{Guo:2012ym,Guo:2012yt,Guo:2011pa}.

Despite the success of QCD sum rules, there remain several subtle issues concerning their application, such as the way to determine the continuum threshold parameter $s_0$ and the uncertainties of the results. In Refs.\cite{Leinweber:1995fn, Benmerrouche:1995qa, Wang:2016sdt, Yuan:2017foa}, a Monte-Carlo-based QCD sum rule is introduced, which starts from the theoretical formula of the SVZ QCD sum rule and uses Monte-Carlo based uncertainty analysis to proceed the study. The Monte-Carlo-based QCD sum rule has some obvious advantages: (a) the continuum threshold $s_0$ can be automatically determined via the fitting procedure; (b) the effects from the uncertainties of the nonperturbative OPE parameters and the phenomenological spectral functions can be more reliably taken into account. Consequently the Monte-Carlo-based QCD sum rule is considered to be able to quantitatively improve the uncertainty analysis,  comparing with the conventional SVZ sum rule, which relies on some empirical ways to obtain the platform of the $s_0$ and Borel transformation parameters.

In this work, we aim at using the both the conventional SVZ and Monte-Carlo-based QCD sum rules in the isoscalar scalar channel to determine the average of the up- and down-quark masses $(m_u+m_d)/2$, which will be shortly denoted as $m_q$ in later discussions. Comparisons from both methods will be analyzed in detail. Another novelty of this work is to implement the sophisticated chiral spectrum functions in the phenomenological side, where one does not need to manually separate the background and the resonance contributions. Furthermore the unitarized chiral spectrum functions are considered to be more suitable for the broad scalar resonances than the Breit-Wigner forms~\cite{Yuan:2017foa}.

The paper is organized as follows. We elaborate the essential formulas of the QCD Sum Rules and the chiral spectrum functions in Sec.II. The numerical analysis and results are given in Sec.III. The last part is left for a short summary and conclusions.

\section{Theoretical formalism}\label{sec-II}

Within the QCD sum rule approach, the connection between quarks/gluons and hadrons is through the two-point correlation function, constructed by two currents, {\it i.e.}
\begin{eqnarray}
  \Pi(q^2) = i \int d^4 x e^{i q \cdot x} \langle \Omega| T\{ j_S(x), j_S^\dagger(0)\}| \Omega \rangle \;,\label{eq:correlator}
\end{eqnarray}
where $j_S$ is the relevant hadronic current, the subscript $S$ represents the scalar current, and $\Omega$ denotes the physical vacuum. In this work, we focus on the current in the isoscalar scalar channel
\begin{eqnarray}\label{eq.scurrent}
 j_S(x) = m_q \frac{1}{\sqrt{2}} \bigg(\bar{u}(x) u(x) + \bar{d}(x) d(x)\bigg) \; ,
\end{eqnarray}
where $m_q = \frac{1}{2} (m_u + m_d)$.

It is stressed that our primary aim in this work is to determine the light-quark mass by using the tool of the QCD sum rule, instead of scrutinizing the internal structures of the scalar resonances. For the latter purpose, it can be crucial to delicately choose the proper currents in the sum-rule analyses. For example, the $\sigma$ resonance has been extensively studied in the QCD sum rule approach by using the mixed currents of the $\bar{q}q$ type and the gluons~\cite{Narison:1988ts,Narison:2000dh}, and also the various tetraquark currents~\cite{Jaffe:1976ig,Jaffe:2004ph,Achasov:1980gu,Latorre:1985uy,Narison:2005wc,Brito:2004tv,Wang:2005xy,Lee:2006vk,Chen:2007zzg,Chen:2007xr}. However, for the purpose to extract the light-quark mass, it is adequate to analyze the $j_S$ current of Eq.~\eqref{eq.scurrent} within the framework of QCD sum rule, as long as one can properly calculate the hadronic contributions to two-point correlation functions of the current, which will be discussed in detail in the following.

The correlation function has two equivalent representations: the QCD side and the phenomenological one. The QCD representation of this function can be calculated by using the Operator Product Expansion (OPE) method. In addition, it has been noticed that the instanton effects could be important in the pseudoscalar and scalar channels~\cite{Shuryak:1982qx,Cherry:2001cj,Elias:1998fs,Elias:1998bq,Shi:1999hm,Yuan:2017foa}. Therefore we also include the instanton contributions in the QCD side, which are calculated within the instanton liquid model~\cite{Shuryak:1982qx,Elias:1998fs}. The correlation function from the QCD side then includes both the OPE and instanton parts
\begin{eqnarray}
  \Pi^{\textrm{(QCD)}} (q^2) = \Pi^{\textrm{(OPE)}}(q^2) + \Pi^{ \textrm{(Inst)} }(q^2) \;.\label{QCD-1}
\end{eqnarray}
After performing the Borel transformation of Eq.(\ref{QCD-1}), one can obtain~\cite{Shuryak:1982qx,Cherry:2001cj,Elias:1998fs,Elias:1998bq,Shi:1999hm,Yuan:2017foa},
\begin{equation}
\label{QCD-2}
\begin{split}
R^{\textrm{(QCD)}}(\tau,\hat m_q)=&R^{\textrm{(OPE)}}(\tau,\hat m_q)+R^{\textrm{(Inst)}}(\tau,\hat m_q)\\
=&m_q^2(1/\sqrt \tau)
\left\{ \frac{3}{8\pi^2} \frac{1}{\tau^2}
\left[ 1 + 4.821098 \frac{\alpha_s(1/\tau)}{\pi} + 21.97646 \left( \frac{\alpha_s(1/\tau)}{\pi} \right)^{2}\right.\right.\\
& + 53.14197
\left. \left( \frac{\alpha_s(1/\tau)}{\pi} \right)^{3} \right] + \frac{\langle\alpha_s G^2\rangle}{8\pi}
\left( 1 + \frac{11}{2}\frac{\alpha_s(1/\tau)}{\pi} \right)\\
&+ 3 \langle m_q \bar qq \rangle \left( 1 + \frac{13}{3} \frac{\alpha_s(1/\tau)}{\pi} \right) - \frac{176}{27}
\pi \kappa \alpha_s\langle \bar qq\rangle^2 \left[ \frac{\alpha_s(1/\tau)}{\alpha_s(\mu^2_0)} \right]^{1/9}  \tau \\
& \left.+ \frac{3}{8\pi^2} \frac{e^{\frac{-\rho^2}{2\tau}}\rho^2}{\tau^3} \left[ K_0 \left( \frac{\rho^2}{2 \tau} \right)
+ K_1 \left( \frac{\rho^2}{2\tau}\right) \right] \right\},
\end{split}
\end{equation}
where the last two terms with the modified Bessel functions $K_0$ and $K_1$ stand for the contributions from the instanton, and $\rho$ represents the instanton size in the instanton liquid model~\cite{Shuryak:1982qx,Elias:1998fs}. In Eq.~\eqref{QCD-2}, $\tau$ is the Borel parameter, $\kappa$ denotes the vacuum factorization violation factor that parameterizes the deviation of the four-quark condensate from a product of two-quark condensates, and $\mu_0$ is the renormalization scale for condensates. The running coupling constant and running mass have been calculated up to four-loop order in the perturbative corrections~\cite{Chetyrkin:1996xa, Shi:1999hm}. The four-loop result with three-flavor quarks for the running coupling constant at scale $1/\sqrt{\tau}$ takes the form~\cite{Chetyrkin:1996xa, Shi:1999hm}
\begin{eqnarray}
\label{eq.alphas}
  \alpha_s(1/\tau) = \pi \left\{\frac{4}{9 } \frac{1}{L} - \frac{256 \ln L}{729L^{2}} + \frac{1}{L^3} \left[ \frac{ 16384 \ln^{2} L }{59049} - \frac{16384 \ln L}{59049} + \frac{6794}{59049}\right] \right\} \; ,
\end{eqnarray}
with $L = \ln \left( \frac{1}{\tau \Lambda_{\text{QCD}}^2}\right)$. The QCD scale is taken  as $\Lambda_{\text{QCD}}=0.353 \,\text{GeV}$ throughout this work~\cite{Narison:2009vy}. The explicit scale $1/\sqrt \tau$ dependence of the light-quark mass in Eq.(\ref{QCD-2}) up to four-loop order is~\cite{Chetyrkin:1996xa, Shi:1999hm}
\begin{eqnarray}
\label{mq-mqhat}
\begin{split}
m_q(1/\sqrt \tau) & = \hat{m}_q \frac{1}{\left(\frac{1}{2} L\right)^{4/9}}  \left\{ 1 + \frac{290}{729}\frac{1}{L} - \frac{256}{729}\frac{\ln L}{L} + \left(\frac{550435}{1062882} - \frac{80\zeta(3)}{729} \right) \frac{1}{L^2} \right. \\
& - \frac{388736}{531441}\frac{\ln L}{L^2} + \frac{106496 }{531441}\frac{\ln^2 L}{L^2} + \left( \frac{2121723161}{2324522934} + \frac{8}{6561}\pi^{4} - \frac{119840 }{531441}\zeta(3) \right. \\
& \left. - \frac{8000 }{59049} \zeta(5) \right)\frac{1}{L^3} + \left(- \frac{611418176}{387420489} + \frac{112640}{531441} \zeta(3) \right) \frac{\ln L}{L^3} \\
&\left. + \frac{335011840}{387420489} \frac{\ln^2 L}{L^3} - \frac{149946386}{1162261467} \frac{\ln^3 L}{L^3} \right\}\,,
\end{split}
\end{eqnarray}
where $\hat m_q$ is the renormalization-group (RG) invariant light-quark mass and $\zeta$ stands for the Riemann zeta function.

On the phenomenological side, we need to provide the phenomenological spectral function which is related to the correlation function through the dispersive integral. It is typical to write the phenomenological spectral function as
\begin{equation}
\label{eq:model1}
 {\rm Im}\Pi^{\textrm{(Phen)}}(s)= {\rm Im}\Pi^{\textrm{(Res)}}(s)
+ {\rm Im}\Pi^{\textrm{(ESC)}}(s)\,,
\end{equation}
where in the conventional QCD sum rule the first term in the right-hand side usually includes the resonance contributions via the Breit-Wigner or $\delta$-function forms, and the second term corresponds to the excited states and continuum. The ${\rm Im}\Pi^{\textrm{(ESC)}}(s)$ part incorporates the excited states and higher energy continuum above the energy scale $s_0$, that should be determined via the sum-rule method.
In this work we use the unitarized chiral form factors~\cite{Guo:2012ym,Guo:2012yt} to construct the resonant part ${\rm Im}\Pi^{\textrm{(Res)}}(s)$ of phenomenological spectral function, which coherently include resonances and the two-meson continuum in the low energy sector.

In the present convention, the spectral function or the imaginary part of the two-point correlator is related to the two-meson form factors through
\begin{eqnarray}
\label{eq:formfactor}
&&{\rm Im}\Pi^{\textrm{(Res)}}(s)=\frac{3 m_\pi^4}{32\pi}\sum_{k=1}^{5} \frac{q_k(s)}{\sqrt{s}}\big|F_k^{\rm{S}}\big|^2\theta(s-s_k^{\rm{threshold}})  \,,
\end{eqnarray}
where the subscript $k$ runs over the relevant two-meson channels: $\pi\pi, K\bar{K}, \eta\eta,\eta\eta'$ and $\eta'\eta'$, $F_{k}^{\rm S}(s)$ denotes the isoscalar-scalar two-meson form factor of the $k$th channel for the $j_S$ current in Eq.~\eqref{eq.scurrent}, and $q_k(s)$ stands for the center of mass three-momentum of the two particles in the $k$th channel. We will keep the same conventions of the form factors as those in Refs.~\cite{Guo:2012ym,Guo:2012yt} and accordingly introduce other relevant coefficients in  Eq.~\eqref{eq:formfactor} to match the normalization of the correlator of $j_S$ in Eqs.~\eqref{eq:correlator} and \eqref{QCD-2}. At low energies the intermediate $\pi\pi$ state is the most important hadronic channel in the isoscalar scalar case. In the meanwhile, we also include the $K\bar{K}, \eta\eta,\eta\eta'$ and $\eta'\eta'$ channels in this work. The scalar form factors $F_{PP'}^{\rm S}(s)$ are calculated in the the framework of the unitarized $U(3)$ resonance chiral theory~\cite{Guo:2012ym,Guo:2012yt}. For the sake of completeness, we recap the construction of the chiral phenomenological spectrum function here. The leading order (LO) $U(3)$ chiral Lagrangian is
\begin{eqnarray} \label{eq.lolagrangian}
\mL^{({\rm LO})}=\frac{ F^2}{4}\langle u_\mu u^\mu \rangle+
\frac{F^2}{4}\langle \chi_+ \rangle
+ \frac{F^2}{3}M_0^2 \ln^2{\det u}\,,
\end{eqnarray}
where the first two terms coincide with the conventional $SU(3)$ chiral perturbation theory~\cite{Gasser:1984gg} and the last one parameterizes the effect of the QCD $U_A(1)$ anomaly, giving rise to the LO mass $M_0$ to the singlet $\eta_0$ meson. The definitions of the basic chiral tensors $u_\mu$ and $\chi_+$ can be found in Refs.~\cite{Guo:2011pa,Guo:2012ym,Guo:2012yt} and references therein. Regarding the higher order effects beyond the LO, one could include the higher order local operators~\cite{Gasser:1984gg} or alternately one can also introduce explicitly the heavier degrees of freedom, such as the bare resonance states~\cite{Ecker:1988te}. The latter method is employed to calculate the two-meson scattering amplitudes and the form factors in Refs.~\cite{Guo:2011pa,Guo:2012ym,Guo:2012yt} within the framework of resonance chiral theory~\cite{Ecker:1988te}. Three types of resonances, including scalar ($S$), vector ($V$) and pseudoscalar ($P$) ones, are considered and the relevant interacting Lagrangians read
\begin{align}\label{lagscalar}
\mL_{S}&= c_d\bra S_8 u_\mu u^\mu \ket + c_m \bra S_8 \chi_+ \ket
 + \widetilde{c}_d S_1 \bra u_\mu u^\mu \ket
+ \widetilde{c}_m  S_1 \bra  \chi_+ \ket \,,
\end{align}
\begin{eqnarray}\label{lagvector}
\mL_{V}= \frac{i G_V}{2\sqrt2}\langle V_{\mu\nu}[ u^\mu, u^\nu]\rangle \,,
\end{eqnarray}
\begin{eqnarray}\label{lagpscalar}
\mL_{P} = i d_m \bra P_8 \chi_- \ket + i \widetilde{d}_m  P_1 \bra \chi_- \ket\,.
\end{eqnarray}
Two relevant local operators at the next-to-leading order that cannot receive contributions after integrating the resonances in Eqs.~\eqref{lagscalar}, \eqref{lagvector} and \eqref{lagpscalar},  are also taken into account
\begin{align}\label{laglam}
\mathcal{L}_{\Lambda} & = \frac{F^2\, \Lambda_1}{12}   \partial^\mu X \partial_\mu X  -\frac{F^2\, \Lambda_2}{12} X \bra \chi_- \ket\,,
\quad (X=\ln \det U)\,.
\end{align}
The chiral one-loop calculation with explicit tree-level resonance exchanges in the $U(3)$ resonance chiral theory is completed in Ref.~\cite{Guo:2011pa}. Later on the one-loop calculation of the strangeness conserving scalar and pseudoscalar form factors is also carried out in a similar theoretical framework in Refs.~\cite{Guo:2012ym,Guo:2012yt}. An algebraic approximation of the N/D method is used to perform the unitarization of the scattering amplitudes and the scalar two-meson form factors. Under this assumption, the unitarized scalar form factor takes the form
\begin{eqnarray}\label{defunitarizedF}
\mathcal{F}(s) &=&  \big[ 1 + N(s) \cdot g(s) \big]^{-1}\cdot R(s)\,,
\end{eqnarray}
with
\begin{eqnarray}
 N(s)&=&{T(s)}^{\rm LO+Res+Loop} + T(s)^{\rm LO}\cdot  g(s)\cdot T(s)^{\rm LO}\,, \\
R(s)&=&{F(s)}^{\rm LO+Res+Loop} + N(s)^{\rm LO}  \cdot g(s)\cdot F(s)^{\rm LO}\,.
\end{eqnarray}
Here $T(s)$ and $F(s)$ stand for the $S$-wave two-meson scattering amplitudes and the scalar form factors from the perturbative calculations, respectively. The  superscripts LO, Res, Loop denote the contributions from the leading order, resonance exchanges (also including the $\Lambda_1$ and $\Lambda_2$ terms) and the chiral loops, in order. By construction, the quantities $N(s)$ and $R(s)$ do not contain any right-hand cut. The latter effect is included in the $g(s)$ function, which is given by
\begin{eqnarray}\label{defgfuncionloop}
 g(s) &=& \frac{1}{16\pi^2}\bigg[ a_{SL}(\mu)+\log\frac{m_2^2}{\mu^2}
-x_+\log\frac{x_+-1}{x_+}
-x_- \log\frac{x_--1}{x_-}  \bigg] \,, \\ \label{defxpm}
x_{\pm} &=&\frac{s+m_1^2-m_2^2}{2s}\pm\frac{1}{-2s}\sqrt{ -4 m_1^2 s+(s+m_1^2-m_2^2)^2}\,. \nonumber
\end{eqnarray}
In coupled-channel case, it becomes a diagonal matrix spanned in the scattering channel space and for the $k$th channel one should replace $m_1$ and $m_2$ in Eq.~\eqref{defgfuncionloop} by the masses of the two particles in that channel. The subtraction constants $a_{SL}$ in Eq.~\eqref{defgfuncionloop} are usually free parameters that need to be fixed in the phenomenological fits. In our unitarization procedure, all the unknown parameters appearing in the unitarized scalar form factors also enter in the unitarized scattering amplitudes. As a consequence, the phenomenological fits to the various experimental scattering observables~\cite{Guo:2011pa,Guo:2012ym,Guo:2012yt} enable one to fix all the unknown parameters in the scalar form factors. The uncertainties of the resonance couplings $G_V$, $c_d$, $c_m$, $\tilde{c}_d$, $\tilde{c}_m$, the subtraction constants and also the bare masses of the resonances have been estimated in Refs.~\cite{Guo:2011pa,Guo:2012yt} by using the error bars of the various experimental data, including the phase shifts and inelasticities from the $\pi\pi\to\pi\pi$, $\pi\pi\to K\bar{K}$, $K\pi\to K\pi$ processes with different isospin and angular momentum numbers and also the $\pi\eta$ event distributions. Such uncertainties of the relevant parameters can be also used to make the  errorbar analyses of the form factors and phenomenological spectrum functions, which will be explicitly shown in the next section.
The spectral function used here clearly improves the one in Ref.~\cite{Yuan:2017foa}, with $\sigma$ and $f_0(980)$ included in the Breit-Wigner formalism, which is likely to be improper for the broad resonance. An additional free parameter is introduced as the weight of the two resonances in their spectral function. In Ref.~\cite{Cherry:2001cj}, the phenomenological spectral density for the isoscalar scalar channel is assumed to be solely contributed by the $\pi\pi$ channel. In our study, the heavier-mass channels, such as the $K\bar{K}, \eta\eta$, $\eta\eta'$ and $\eta'\eta'$, are explicitly included within the $U(3)$ chiral framework.

For the ESC contributions in the phenomenological spectral density in Eq.~\eqref{eq:model1}, the traditional ESC model is utilized in this work~\cite{Yuan:2017foa,Shi:1999hm},
\begin{equation}
\label{eq:esc}
\begin{split}
\frac{1}{\pi}{\rm Im} \Pi^{\textrm{(ESC)}}(s) & = m_q^2 (1/\sqrt \tau)
\left\{ \frac{3s}{8\pi^2} \left[ 1 + \frac{\alpha_s}{\pi} \big( \frac{17}{3} - 2 \ln \big( s \tau \big) \big) + \big( \frac{\alpha_s}{\pi}\big)^{2} \big( 31.8640 \right.\right.\\
&- \frac{95}{3} \ln \big( s \tau \big)
+ \frac{17}{4} \ln^{2} \big(s \tau \big) \big) + \big( \frac{\alpha_s}{\pi}\big)^{3} \big( 89.1564 - 297.596 \ln \big( s \tau \big)  \\
&\left.\left. + \frac{229}{2} \ln^{2} \big(s \tau \big) - \frac{221}{24} \ln^{3} \big( s \tau \big) \right] - \frac{3}{4\pi} s J_1(\sqrt{s} \rho) Y_1(\sqrt{s} \rho) \right\}\theta(s - s_0)\,,
\end{split}
\end{equation}
where $J_1$ and $Y_1$ are Bessel functions of the first and second kind, respectively, and $s_0$ is the continuum threshold separating the contributions from higher excited states and continuum. It should be noted that we include in Eq.~\eqref{eq:esc} the perturbative contribution from the quark-hadron duality up to four-loop order~\cite{Shi:1999hm}.

By performing the Borel transformation of Eq.(\ref{eq:model1}), one can obtain the phenomenological representation for the correlation function
\begin{eqnarray}
\label{eq:phen}
R^{\textrm{(Phen)}}(\tau,s_0,\hat m_q) &=& \frac{1}{\pi} \int_{4 m_\pi^2}^\infty e^{-s\tau} {\rm Im}\Pi^{\textrm{(Phen)}}(s) \, ds \nonumber \\ &=& R^{\textrm{(Res)}}(\tau, s_0) + R^{\textrm{(ESC)}} (\tau, s_0, \hat m_q)\,.
\end{eqnarray}
The master equation for QCD sum rules can be established by matching the results from the QCD calculation in Eq.~\eqref{QCD-2} with its counter part of the phenomenological one in Eq.~ \eqref{eq:phen}, $i.e.$,
\begin{equation}
\label{eq:master}
R^{\textrm{(QCD)}}(\tau, \hat m_q) = R^{\textrm{(Phen)}}(\tau, s_0, \hat m_q).
\end{equation}

Within the SVZ QCD sum rules, in order to extract the light quark mass $m_q$, we firstly derive the expression of $\hat{m_q}$ from Eq.(\ref{eq:master}). After an analytical manipulation, we obtain
\begin{eqnarray}
  \hat{m}_q = \frac{m_\pi^2\sqrt{3}}{8\pi} \sqrt{\frac{ \int_{4m_\pi^2}^{s_0} e^{-s \tau} {\rm Im}\Pi^{\textrm{(phen)}}(s) ds}{f^2(\tau) g(\tau)- \int_{s_0}^\infty e^{-s \tau} f^2(\tau) h(\tau, s) ds}}\, ,
\end{eqnarray}
where the functions $f(\tau), \, g(\tau)$, and $h(\tau, s)$ are defined as
\begin{eqnarray}
\begin{split}
 f(\tau) \equiv& \frac{1}{\left(\frac{1}{2} L\right)^{4/9}}  \left\{ 1 + \frac{290}{729}\frac{1}{L} - \frac{256}{729}\frac{\ln L}{L} + \left(\frac{550435}{1062882} - \frac{80\zeta(3)}{729} \right) \frac{1}{L^2} \right. \\
& - \frac{388736}{531441}\frac{\ln L}{L^2} + \frac{106496 }{531441}\frac{\ln^2 L}{L^2} + \left( \frac{2121723161}{2324522934} + \frac{8}{6561}\pi^{4} - \frac{119840 }{531441}\zeta(3) \right. \\
& \left. - \frac{8000 }{59049} \zeta(5) \right)\frac{1}{L^3} + \left(- \frac{611418176}{387420489} + \frac{112640}{531441} \zeta(3) \right) \frac{\ln L}{L^3} \\
&\left. + \frac{335011840}{387420489} \frac{\ln^2 L}{L^3} - \frac{149946386}{1162261467} \frac{\ln^3 L}{L^3} \right\} \,,
\end{split}
\end{eqnarray}
\begin{eqnarray}
\begin{split}
g(\tau) \equiv & \frac{3}{8\pi^2} \frac{1}{\tau^2} \left\{\left[ 1 + 4.821098 \frac{\alpha_s (1/\tau)}{\pi} + 21.97646 \left( \frac{\alpha_s(1/\tau)}{\pi} \right)^{2} + 53.14197 \left( \frac{\alpha_s(1/\tau)}{\pi}\right)^{3}  \right] \right.\\
& + \frac{\langle \alpha_s G^2 \rangle}{8\pi}
\left( 1 + \frac{11}{2} \frac{\alpha_s(1/\tau)}{\pi}\right)
+ 3\langle m_q \bar qq \rangle \left( 1 + \frac{13}{3} \frac{\alpha_s(1/\tau)}{\pi}\right) \\
&\left. - \frac{176}{27}
\pi \kappa \alpha_s \langle \bar qq \rangle^2 \left[ \frac{\alpha_s(1/\tau)}{\alpha_s(\mu^2_0)} \right]^{1/9} \tau + \frac{3}{8\pi^2}\frac{e^{\frac{-\rho^2}{2\tau}}\rho^2}{\tau^3} \left[ K_0 \left( \frac{\rho^2}{2 \tau} \right)
+ K_1 \left( \frac{\rho^2}{2\tau}\right)\right] \right\}\,,
\end{split}
\end{eqnarray}
\begin{eqnarray}
\begin{split}
h(\tau, s) \equiv & \left\{ \frac{3s}{8\pi^2} \left[ 1 + \frac{\alpha_s(1/\tau)}{\pi} \big(\frac{17}{3} - 2 \ln \big( s \tau \big) \big) + \big( \frac{\alpha_s(1/\tau)}{\pi}\big)^{2} \big( 31.8640 - \frac{95}{3} \ln \big( s \tau \big)\right.\right.\\
& + \frac{17}{4} \ln^{2} \big( s \tau \big) \big) + \big(\frac{\alpha_s(1/\tau)}{\pi}\big)^{3} \big(89.1564 - 297.596 \ln \big( s \tau \big)\\
&\left.\left. + \frac{229}{2} \ln^{2} \big( s \tau \big) - \frac{221}{24} \ln^{3} \big( s \tau \big) \big) \right] - \frac{3}{4\pi} s J_1(\sqrt{s} \rho) Y_1(\sqrt{s} \rho) \right\} \theta(s-s_0)\,.
\end{split}
\end{eqnarray}
Then we can obtain the scale-running light-quark mass $m_q (1/\sqrt{\tau})$ from Eq.(\ref{mq-mqhat}).

In this work, we will follow another version of the QCD sum rules, that is the so-called Monte-Carlo-based QCD sum rules~\cite{Leinweber:1995fn, Wang:2011zzx}, to determine the threshold parameter $s_0$ and the RG-invariant light quark mass $\hat{m}_q$. To be more specific, the values of $s_0$ and $\hat{m}_q$ will be determined by minimizing the following quantity
\begin{equation}
\label{eq:chi2}
\chi^2_{k,\xi}=\sum_{j=1}^{n_B}\frac{[R^{\textrm{(QCD)}}_k(\tau_j,\hat m_{q}^{k,\xi})
-R^{\textrm{(phen)}}_\xi(\tau_j,s_{0}^{k,\xi}, \hat m_{q}^{k,\xi})]^2}{\sigma_{\text{QCD}}^2(\tau_j)}\,.
\end{equation}
In order to quantitatively estimate the error bars of the resulting $s_0$ and $\hat{m}_q$ caused by the uncertainties from the input QCD parameters, we generate large amount of $R^{\textrm{(QCD)}}_k$ by randomly varying the OPE condensate parameters and the instanton size $\rho$ within ten percent uncertainties and randomly sampling the $\kappa$ parameter in the range $2 \sim 4$. Meanwhile, we also use the results from Ref.~\cite{Guo:2012yt} to randomly generate large samples of phenomenological spectral functions. The index $k$ in Eq.~\eqref{eq:chi2} corresponds to the $k$th configuration among the randomly generated QCD-parameter samples, the index $\xi$ runs over different samples of the phenomenological spectral functions. The Borel parameter $\tau_j = \tau_\textrm{min} + (\tau_\textrm{max} - \tau_\textrm{min}) \times (j-1)/(n_B-1)$, runs over the appropriate window $[\tau_{\text{min}}, \tau_{\text{max}}]$, which range will be explained later in next section. We will take $n_B=21$ in Eq.~\eqref{eq:chi2} as in Ref.~\cite{Leinweber:1995fn, Wang:2011zzx}. In fact, we have tried to increase the values of $n_B$ in the calculation, and it turns out that larger values of $n_B$ barely affect the results.
The large samples from the randomly generated $R^{\textrm{(QCD)}}_k$ and $R^{\textrm{(phen)}}_\xi$ will then be implemented in Eq.~\eqref{eq:chi2} to repeatedly determine the $s_0$ and $\hat{m}_q$.  The large amount of the fitted $s_0$ and $\hat{m}_q$ will be used for  uncertainty studies.

\begin{figure}[hbpt]
\begin{center}
\includegraphics[width=0.95\textwidth]{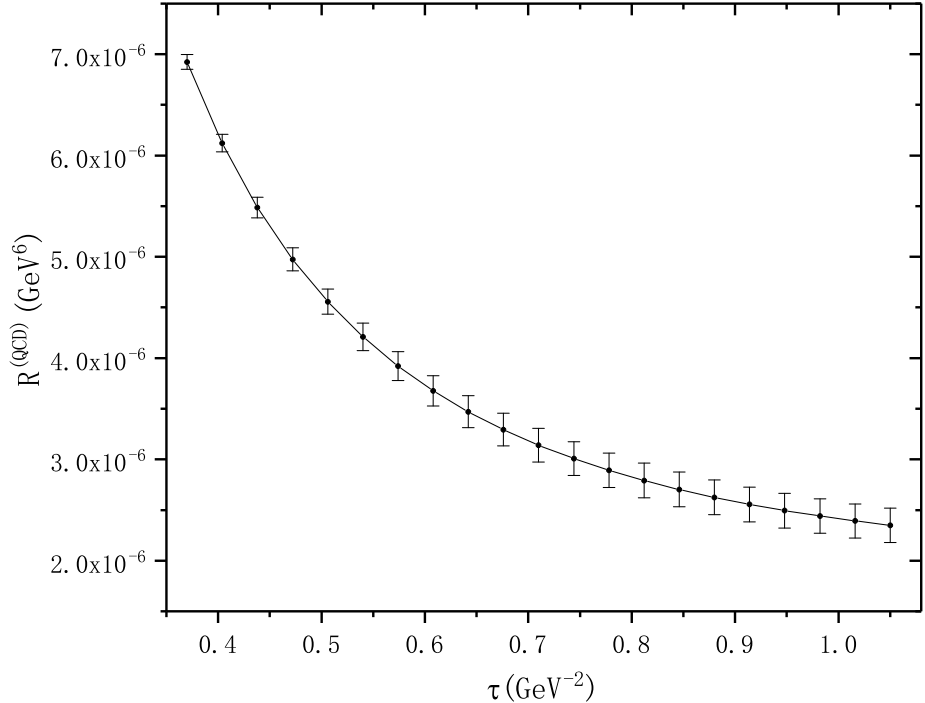}
\caption{ $R^{\text{(QCD)}}$ in Eq.~\eqref{QCD-2}, together with its uncertainties $\sigma_{\rm QCD}(\tau_{j=1,2,...,21})$ in Eq.~\eqref{eq.sigmaqcd}, as functions of the Borel parameter $\tau$. See the text for details about the estimation of the uncertainty $\sigma_{\rm QCD}$.   }
\label{fig.rqcdsigma}
\end{center}
\end{figure}

The quantity $\sigma_{\text{QCD}}^2$ in Eq.~\eqref{eq:chi2} corresponds to the variance of the $R^{\rm{(QCD)}}_{k=1,2,...n_S}$. In our study we generate a large sample of $R^{\rm{(QCD)}}_{k=1,2,...n_S}$ with $n_S=20000$ at each Borel parameter $\tau_j$. At the $j$th Borel parameter $\tau_j$ the variance $\sigma_{\text{QCD}}^2$ takes the standard definition
\begin{eqnarray}\label{eq.sigmaqcd}
\sigma_{\text{QCD}}^2 (\tau_j) = \frac{1}{(n_S - 1)} \sum_{k=1}^{n_S} [R_k^{\text{(QCD)}} (\tau_j)-\overline{R^{\text{(QCD)}}}(\tau_j)]^2,
\end{eqnarray}
where $k$ denotes the $k$th set of the randomly generated $n_S$ QCD-parameter samples, and $\overline{R^{\text{(QCD)}}}(\tau_j)$ corresponds to the mean value of the huge samples $R^{\rm{(QCD)}}_{k=1,2,...n_S}$. It is pointed out that we estimate the values of  $\sigma_{\text{QCD}}^2(\tau_j)$ by taking into account the explicit dependences of $R^{\textrm{(QCD)}}$ on the OPE condensates, the instanton size $\rho$, and the $\kappa$ in Eq.~\eqref{QCD-2}. Since the quantity $\hat{m}_q$ is implicitly dependent on the condensates and the parameters $\rho$ and $\kappa$, we do not explicitly take the uncertainties of $\hat{m}_q$ when estimating $\sigma_{\text{QCD}}^2$ and $\hat{m}_q$ will be fixed at the value obtained by taking the central values of the relevant input QCD parameters. To be more specific, we firstly calculate the random configurations of the $R^{\text{(QCD)}}$ by assigning ten-percent uncertainties to the OPE condensate parameters and the instanton size $\rho$, and also generating random samples of $\kappa$ between 2 and 4. All the random samples of the input parameters are generated with Gaussian distributions. Afterwards, we evaluate the standard variance $\sigma_{\text{QCD}}^2$ of the large data sets of $R^{\text{(QCD)}}$. Next, we use the $\sigma_{\text{QCD}}^2$ in the expression of $\chi^2$ to fit the optimal results $s_0$ and $\hat{m}_q$. The resulting curves of $R^{\text{(QCD)}}(\tau)$, together with the uncertainties $\sigma_{\rm QCD}(\tau_{j=1,2,...,21})$, are explicitly shown in Fig.~\ref{fig.rqcdsigma}.  

\section{Numerical results}

For the numerical analysis, we follow Ref.~\cite{Yuan:2017foa} to implement ten-percent uncertainties for the OPE parameters evaluated at $\mu_0=1\;\text{GeV}$, whose explicit values read
\begin{eqnarray}
\begin{aligned}
& \langle \alpha_s G^2\rangle = (0.070 \pm 0.007) \, \text{GeV}^4 \; ,\;
\langle m_q \bar{q} q \rangle = - (0.10 \pm 0.01 ) \, \text{GeV}^4 \; ,\\
& \kappa \alpha_s \langle \bar{q} q \rangle^2 = (1.49\pm 0.15)\times 10^{-4} \kappa \; \text{GeV}^6 \; ,\;
\rho = (1.67\pm 0.17) \; \text{GeV}^{-1} \; ,
\end{aligned}
\label{QCD-input-parameters}
\end{eqnarray}
where the factor $\kappa$ is introduced to quantify the violation of factorization hypothesis in estimating the dimension-6 quark condensates. The values of $\kappa$ have been determined in the range of $2 \sim 4$ in different channels~\cite{Narison:2009vy,Chung:1984gr,Narison:1995jr}. The influences of the $\kappa$ parameter and the instanton on the OPE expansion, will be discussed in a moment.

\begin{figure}[hbpt]
\begin{center}
\includegraphics[width=0.95\textwidth]{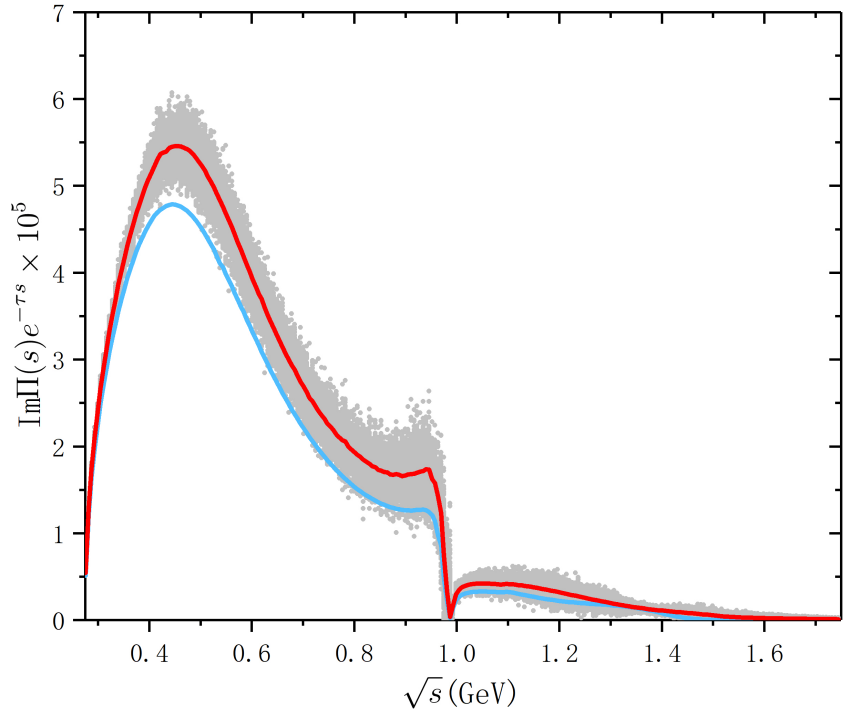}
\caption{The phenomenological spectrum functions ${\rm Im}\Pi(s)$ multiplied by the Borel transformation factor $e^{-\tau s}$ with $\tau=1.0$~GeV$^{-2}$ as a function of energy. The red line corresponds to the spectral function by using the best fit parameters and the corresponding amplitudes given in Ref.~\cite{Guo:2012yt} and the surrounding gray areas are randomly generated according to the uncertainties estimated in the same reference. The blue line stands for the spectral function by taking the best fit parameters and related amplitudes from Ref.~\cite{Guo:2011pa}. In both cases, the effects in the energy region beyond 1~GeV are found to be small. }
\label{exp-spectrum}
\end{center}
\end{figure}

For the input phenomenological spectral functions, we will take the analyses from Refs.~\cite{Guo:2012yt,Guo:2011pa}. In the former reference two different multiplets of bare scalar resonances and one multiplet of pseudoscalar resonances are included in the study of the spectral functions, while in the latter reference only one multiplet of bare scalar states are considered. As a result, somewhat different values of the parameters are also obtained in Refs.~\cite{Guo:2011pa,Guo:2012yt}. In the following discussions, the values obtained by using  the phenomenological spectral functions of Ref.~\cite{Guo:2012yt} will be considered as the preferred central results. The deviations resulting from the use of spectral functions of Ref.~\cite{Guo:2011pa} will be considered as systematic uncertainties to the central results. In addition, we also randomly generate large samples of phenomenological spectral functions according to the results from Ref.~\cite{Guo:2012yt} (see also the discussions in the last section), which will be used to estimate the statistical uncertainties of the $s_0$ and $\hat{m}_q$. The phenomenological spectral functions after including the Borel transformation factor $e^{-\tau s}$ are explicitly shown in  Fig.~\ref{exp-spectrum}. It is obvious that the contributions in the energy region below 1~GeV are dominant and the effects from the high energy tails beyond the 1.4~GeV are tiny and negligible.

\subsection{ Effects of the parameter $\kappa$ and the instanton in $R^{\text{(QCD)}}$}

The sizes of the dimension-six four-quark condensates have been intensively discussed in the literature~\cite{Narison:2009vy,Chung:1984gr,Narison:1995jr,Gimenez:1989bm,Kremer:1984bg,Braaten:1991qm,
Bertlmann:1984ih,Launer:1983ib,Chetyrkin:1998yr,Cherry:2001cj,Dominguez:1984yx,Wang:2016sdt,Wang:2021dcb}, and there is still no consensus on their precise values. It is plausible that the magnitudes of the four-quark condensates are underestimated in the vacuum saturation assumption~\cite{Shifman}, meaning that the $\kappa$ parameter can be obviously larger than one. Indeed a wide range of $\kappa$ between 2 and 4~\cite{Chung:1984gr,Narison:2009vy,Narison:1995jr,Gimenez:1989bm,Cherry:2001cj,Wang:2016sdt}, is determined from various phenomenological processes, including the $e^+ e^-\to hadrons$, the $\tau \to hadron$ decays, the baryon spectra, etc. A recent analysis by combing the electron-positron annihilation and the inclusive $\tau$ decay processes gives the preferred value of $\kappa =3.0 \pm 0.2$~\cite{Narison:2009vy}, which fits well in the wide range $2 \sim 4$. In this work, we make a very conservative estimate of the $\kappa$ by taking its value as $3.0\pm 1.0$, and assess its influence on the determination of the light-quark mass.

\begin{figure}[hbpt]
\begin{center}
\includegraphics[width=0.95\textwidth]{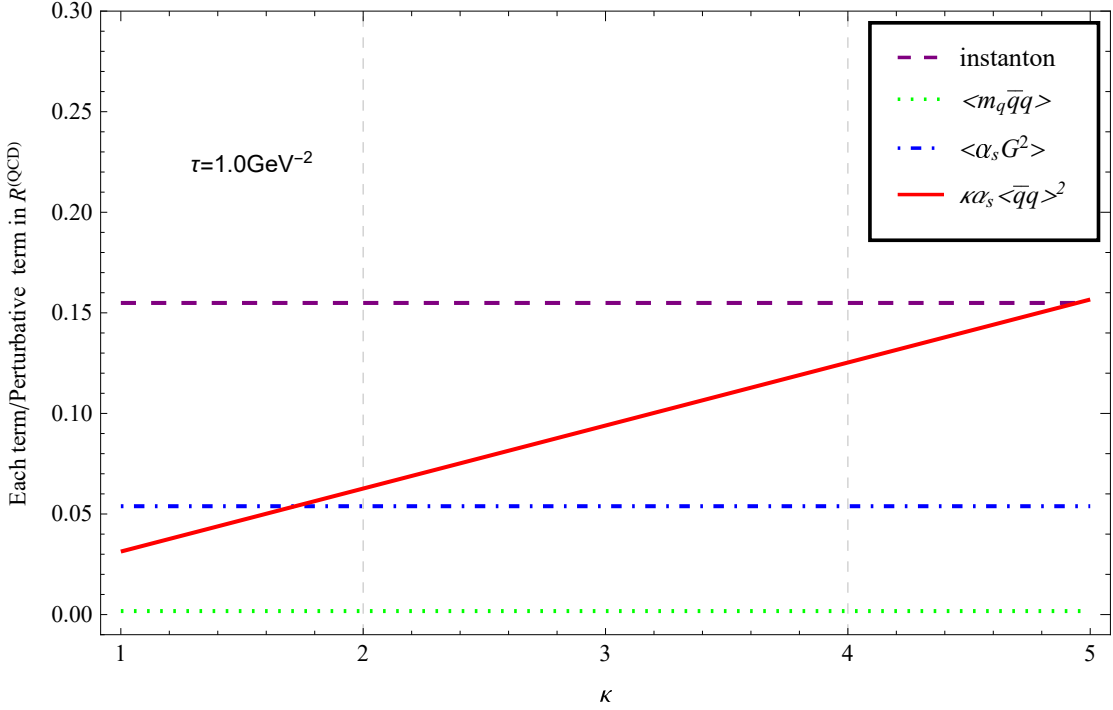}
\caption{The ratios of the various nonperturbative contributions with respect to the perturbative QCD part in $R^{\text{(QCD)}}$. The two vertical dashed lines represent the conservative range of the parameter $\kappa$ used in our study. }
\label{ratio-kappa}
\end{center}
\end{figure}

In order to clearly show the effects of the $\kappa$, the instanton and other nonperturbative terms in the QCD spectral function~\eqref{QCD-2}, we give in Fig.~\ref{ratio-kappa} the ratios of the various contributions with respect to the perturbative QCD result by taking $\tau=1.0\, \text{GeV}^{-2}$. It should be noted that the contribution of the four-quark condensate is negative and we take its absolute value in Fig.~\ref{ratio-kappa} in order to clearly show the sizes of different nonperturbative contributions. 

\begin{figure}[hbpt]
\begin{center}
\includegraphics[width=0.95\textwidth]{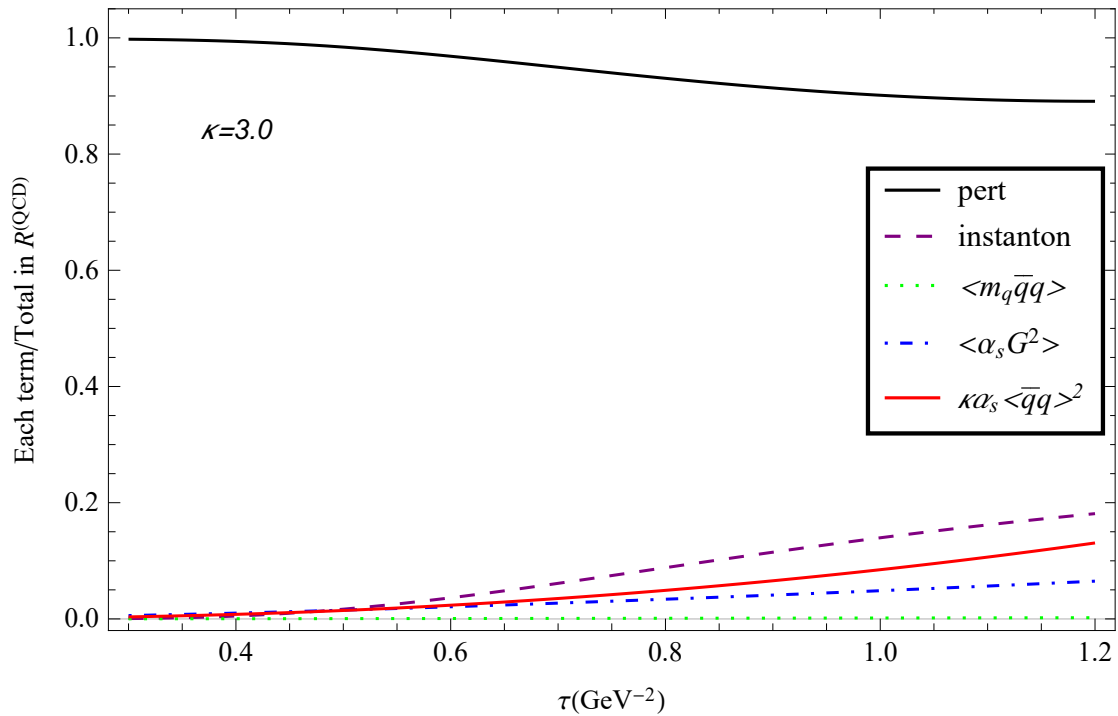}
\caption{The $\tau$ dependence of the ratios of various terms with respect to the full $R^{\text{(QCD)}}$ in Eq.~\eqref{QCD-2}. The results by taking $\kappa=3.0$ are explicitly given here. The curves by taking different values of $\kappa$ between 2.0 and 4.0 are quantitatively similar, and hence are not explicitly shown. }
\label{ratio-tau}
\end{center}
\end{figure}

According to Fig.~\ref{ratio-kappa}, it is obvious that the magnitude of the four-quark condensate is larger than the two-quark and the two-gluon condensates, but is smaller than the instanton contribution for $2<\kappa<4$. It is clear that the contribution from the perturbative QCD term dominates $R^{\text{(QCD)}}$ in Eq.~\eqref{QCD-2}. Furthermore, we also study the dependence of the various terms in the OPE with the Borel parameter $\tau$, and the results are given in Fig.~\ref{ratio-tau}. To be definite, we show the curves by taking $\kappa=3.0$ and verify that the results by taking different values of $\kappa$ are quantitatively similar. As in Fig.~\ref{ratio-kappa}, the perturbative QCD term is the dominant part, while the contributions from the nonperturbative parts, including the instanton, and the quark and gluon condensates, slightly vary when the Borel parameter $\tau$ is increased up to 1.2~GeV$^{-2}$. The sensitivity of the parameter $\kappa$ is related to the relative size of the four-quark condensate to the perturbative QCD term, which will give part of the uncertainties of the Borel window $\tau$, $s_0$ and $\hat{m}_q$ in our analysis. Detailed studies on the error-bar analyses will be given in the next section.

\subsection{Results from the conventional SVZ sum rules}

In the conventional SVZ QCD sum rules, the parameters $\tau$ and $s_0$ are fixed according to two well accepted criteria~\cite{Shifman, Reinders:1984sr, P.Col}. The first one is required by the convergence of the OPE. One needs to find a proper region for $\tau$ to guarantee that the contribution from the highest dimension operator (HDO) included is less than $10\%$ in the OPE side. The degree of convergence can be defined as
\begin{eqnarray}\label{eq.ratioope}
  \text{ratio}_{\langle \bar{q} q\rangle^2} = \frac{R^{\text{(OPE)}}_{\langle \bar{q} q\rangle^2}(\tau, \hat{m}_q)}{R^{\text{(OPE)}}(\tau, \hat{m}_q)}\,,
\end{eqnarray}
where $R^{\text{(OPE)}}(\tau, \hat{m}_q)$ represents the overall OPE contributions in Eq.(\ref{QCD-2}), and the subscript $\langle \bar{q} q \rangle^2$ refers the HDO condensate here.

The second criterion requires that the portion of resonance contribution (RC), $i.e.$, the contributions from the energy region below $s_0$, should be over 50\%~\cite{P.Col}, which can be formulated as
\begin{eqnarray}\label{eq.ratiorc}
  \text{ratio}^{\text{RC}} = \frac{R^{\text{(QCD)}}(\tau,\hat{m}_q,s_0 )}{R^{\text{(QCD)}}(\tau, \hat{m}_q)} \,, \label{RatioRC}
\end{eqnarray}
being $R^{\text{(QCD)}}(\tau,\hat{m}_q,s_0 )$ the effects from the QCD side below $s_0$. Under this prerequisite, the contributions of higher excited and continuum states will be suppressed.

\begin{figure}[hbpt]
\begin{center}
\includegraphics[width=0.44\textwidth]{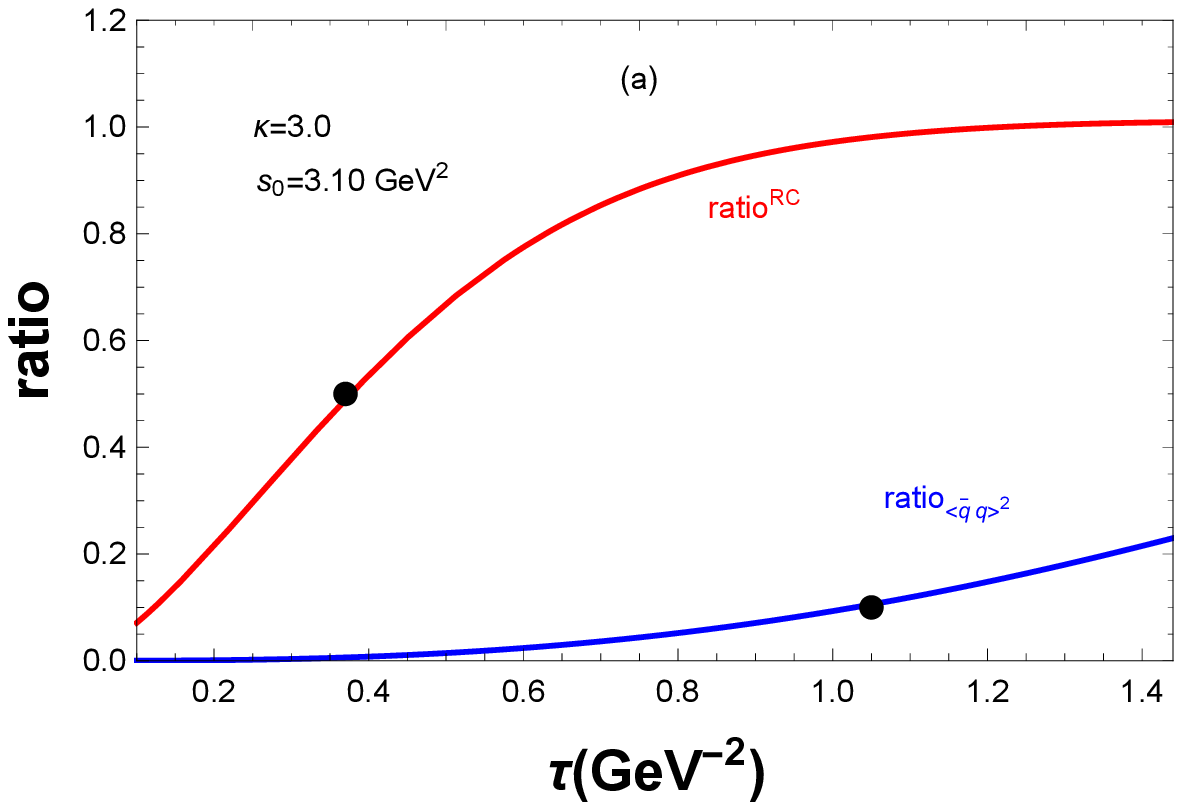}
\includegraphics[width=0.44\textwidth]{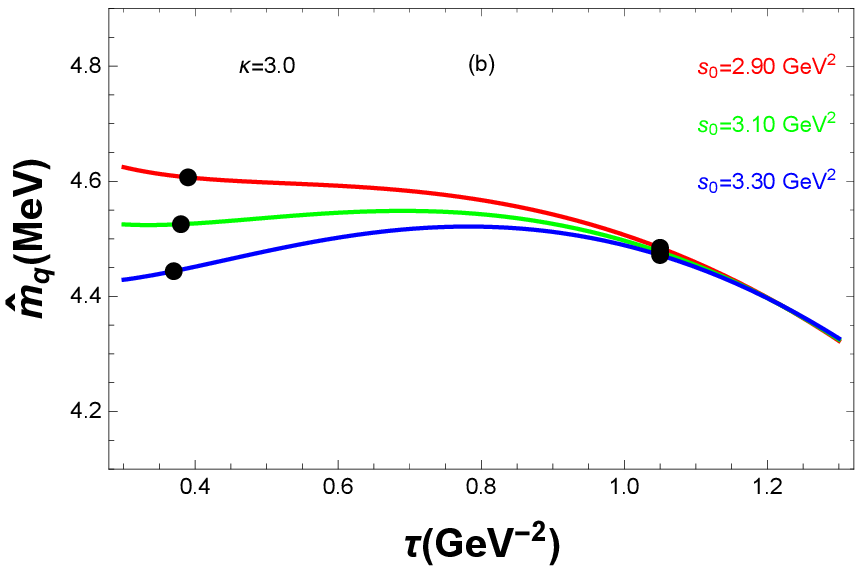}
\caption{(a) The OPE convergence $\text{ratio}_{\langle \bar{q} q \rangle^2}$ and the resonance contribution $\text{ratio}^{\text{RC}}$ as functions of the Borel parameter $\tau$ with $s_0 = 3.10\; \text{GeV}^2$ and $\kappa = 3.0$. The blue and red lines represent $\text{ratio}_{\langle \bar{q} q \rangle^2}$ and $\text{ratio}^{\text{RC}}$, respectively. (b) The RG-invariant light-quark mass $\hat{m}_q$ as a function of the Borel parameter $\tau$ at different values of $s_0$ with $\kappa = 3.0$. The black dots on each line indicate the lower/upper limits of the valid Borel window. }
\label{ratio-mass-25}
\end{center}
\end{figure}

The standard procedure of the SVZ sum rules is to mutually vary the $s_0$ and $\tau$ to find their proper values, in order to satisfy the two aforementioned criteria and find a smooth plateau for the $\hat{m}_q$. On the smooth plateau, the RG-invariant light quark mass $\hat{m}_q$ should be in principle independent of the Borel parameter $\tau$, or at least only shows weak dependence. During this procedure, the preferred value of $s_0$ is found to be 3.10~GeV$^2$ for $\kappa = 3.0$.  We will tentatively assign a several-percent uncertainty for the $s_0$ as $3.10\pm 0.20$~GeV$^2$ in later discussions~\cite{Qiao:2013raa}. The fulfillment of the two criteria and smooth plateaus are shown in Fig.~\ref{ratio-mass-25}. The dependences of two ratios of Eqs.~\eqref{eq.ratioope} and \eqref{eq.ratiorc} with the Borel parameter $\tau$ at the preferred $s_0 = 3.10 $~GeV$^2$ are given in the left panel, from which one can easily identify the valid region of $\tau$. In the right panel of Fig.~\ref{ratio-mass-25}, we show the rather weak dependence of $\hat{m}_q$ with the Borel parameter $\tau$.
By taking into account of the error bars of the OPE condensates, the instanton size and the $\kappa$ parameter, we repeatedly take the standard procedure to determine the values of $s_0$ and $\tau$. The results of the $s_0$ and $\tau$ with their uncertainties extracted from the SVZ sum rules are summarized in Table~\ref{SVZ-results}.

\begin{table}
\caption{Results from the SVZ QCD sum rules. The uncertainties of the Borel windows $\tau_{\rm min,max}$, $s_0$ and $\hat{m}_q$ are obtained by taking into account the error bars of the OPE parameters in Eq.~\eqref{QCD-input-parameters} and $\kappa=3.0\pm 1.0$.   }
\begin{center}
\renewcommand\arraystretch{1.1}
\begin{tabular}{|c|c|c|c|c|c|}
\hline\hline    $\tau_{\text{min}}/\text{GeV}^{-2}$ & $\tau_{\text{max}}/\text{GeV}^{-2}$ &  $s_0/\text{GeV}^2$ & $\hat{m}_q /\text{MeV}$ & $m_q$ (2 GeV)/MeV\\
\hline   $0.37\pm 0.02$  & $1.05^{+0.19}_{-0.13}$ & $3.10 \pm 0.20$  & $4.54^{+0.21}_{-0.29} $ & $3.46^{+0.16}_{-0.22}$ \\
\hline\hline
\end{tabular}
\end{center}
\label{SVZ-results}
\end{table}

A slightly different treatment of the phenomenological spectral function from Ref.~\cite{Guo:2011pa}, provides another source of uncertainty to $s_0$ and $m_q$, compared to the results in Table~\ref{SVZ-results} obtained by using the spectral functions from Ref.~\cite{Guo:2012yt}. The parameters by taking the phenomenological spectral functions from Ref.~\cite{Guo:2011pa} with $\kappa=3.0$ are found to be
\begin{eqnarray} \label{eq.svzparamprd11}
s_0= 2.90~{\rm GeV^2}\,, \quad \hat{m}_q= 4.12~{\rm MeV}\,, \quad m_q= 3.13~{\rm MeV}\,.
\end{eqnarray}
The deviations between the results in Eq.~\eqref{eq.svzparamprd11} and the central values in Table~\ref{SVZ-results} are considered as systematic uncertainties caused by the different treatments of spectral functions. In this way, our determination of the RG-invariant light-quark mass $\hat{m}_{q}$ via the SVZ QCD sum rules reads
\begin{eqnarray}
\hat{m}_{q}= 4.54^{+0.21}_{-0.29} \pm 0.42 \, \text{MeV},
\end{eqnarray}
where the first error bar stems from the uncertainties of Borel parameter $\tau$, threshold parameter $s_0$, the condensates, the instanton size $\rho$ and the factor $\kappa$, and the second one is originated by taking two different sets of spectral functions~\cite{Guo:2011pa,Guo:2012yt}. Similarly for the scale dependent light-quark mass, our determination takes the value
\begin{eqnarray}
m_q(2 \, \text{GeV}) = 3.46^{+0.16}_{-0.22} \pm 0.33 \, \text{MeV}\,.
\label{mq-SVZ-results}
\end{eqnarray}

\subsection{Results from the Monte-Carlo-based sum rules}

As an complementary study, we also analyze the RG-invariant light-quark mass via the Monte-Carlo-based QCD sum rule, which is expected to give a more solid quantitative estimation of the uncertainty. In this approach, the values of $s_0$ and $\hat{m}_q$ are obtained by minimizing the $\chi^2$ functions of Eq.~\eqref{eq:chi2} for a given $R^{\rm (QCD)}$ and $R^{\rm (phen)}$. According to the results from the SVZ approach shown in Fig.~\ref{ratio-mass-25}, we will take a conservative range for $0.37\,{\rm GeV}^{-2}<\tau<1.05\,{\rm GeV}^{-2}$ in the Monte-Carlo-based sum rules. We have explicitly verified that the two criteria of the SVZ sum rules in the conservative ranges of $\tau$ with the solutions of $s_0$ from the minimization procedures are generally well satisfied.

\begin{figure}[htbp]
\centering
\includegraphics[width=0.7\textwidth]{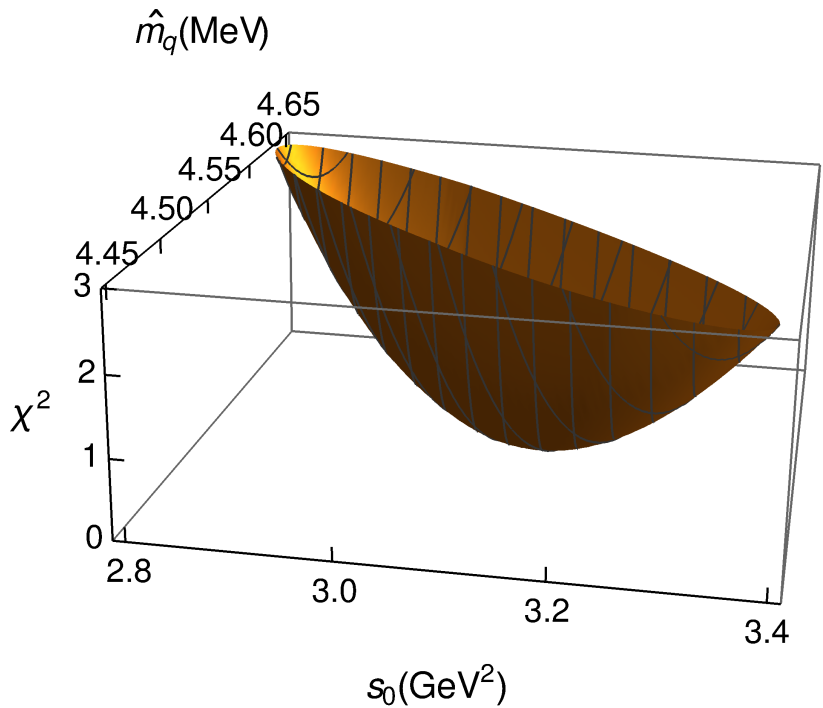}
\caption{$\chi^2$ as a function of $s_0$ and $\hat{m}_q$, where the central condensate values, the central phenomenological spectrum and $\kappa = 3.0$ are taken. A strong correlation between the $s_0$ and $\hat{m}_q$ can be clearly seen. }\label{plot3d}
\end{figure}

\begin{figure}[hbpt]
\begin{center}
\includegraphics[angle=-0,width=0.49\textwidth]{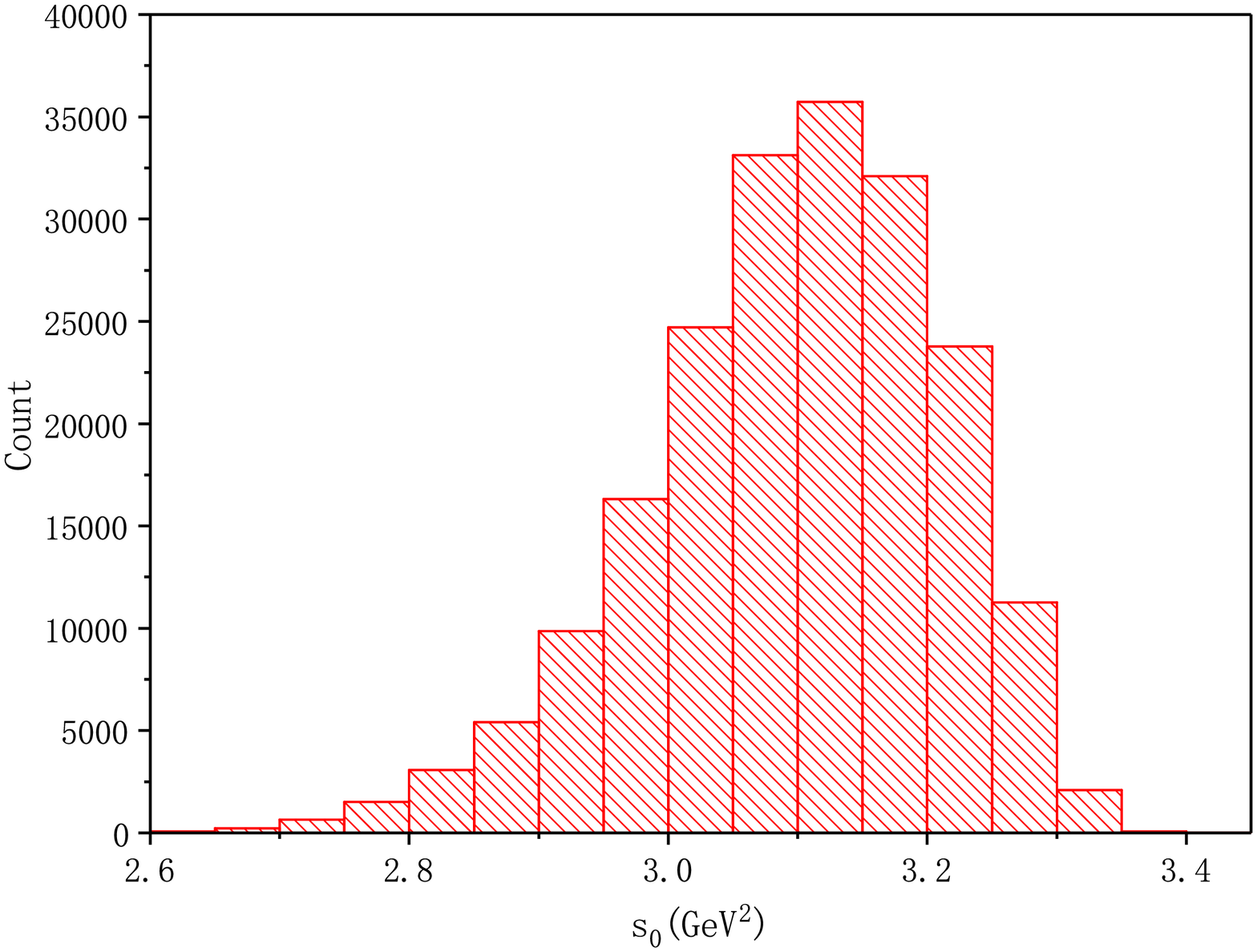}
\includegraphics[angle=-0,width=0.49\textwidth]{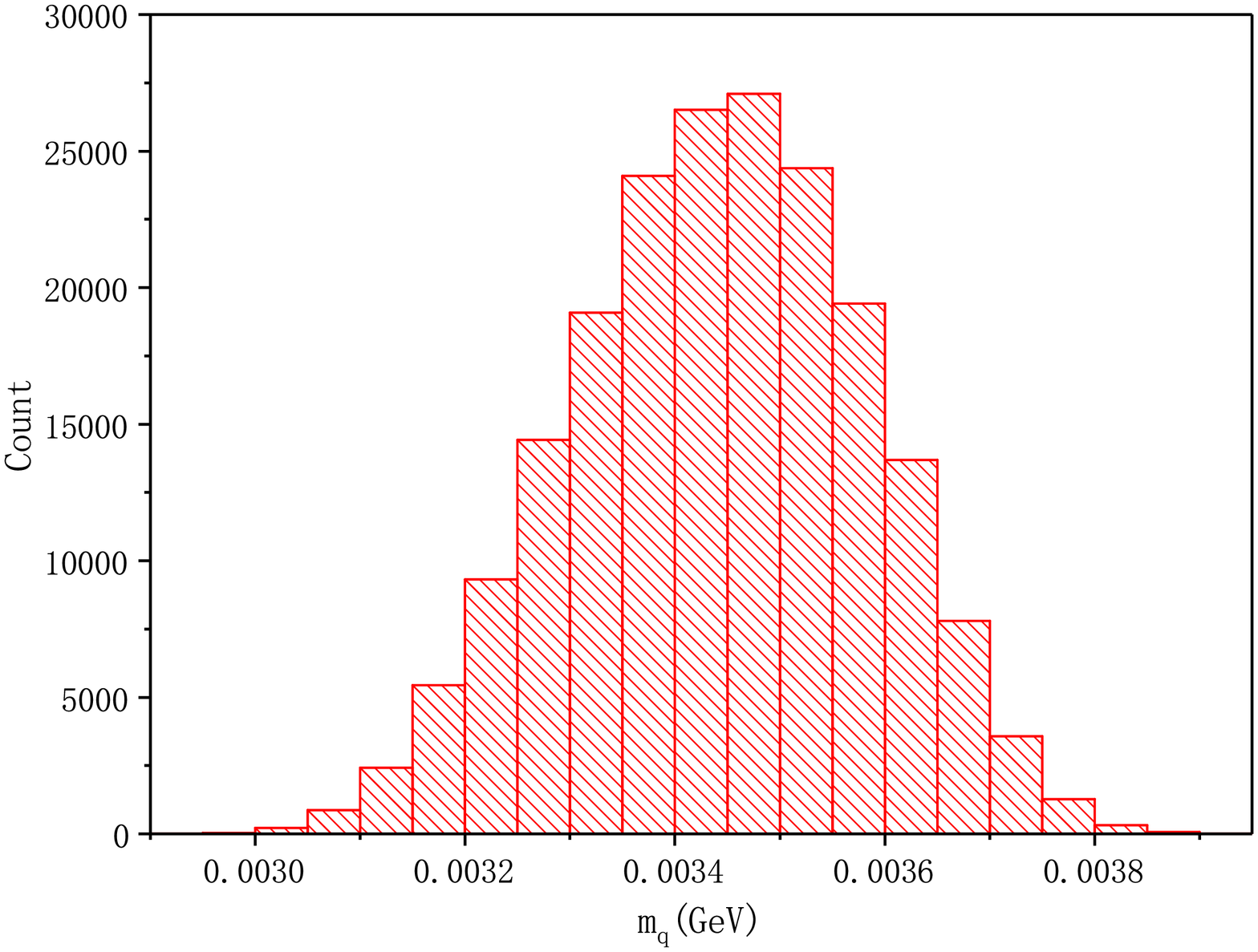}
\caption{The histogram distributions of the resulting parameters $s_0$ and $m_q$ (obtained from the RG-invariant $\hat{m}_q$) from the repeated $\chi^2$ minimization procedures by taking the large amount of random samples of the OPE condensates, the instanton size $\rho$, the parameter $\kappa$ and the spectral functions from Ref.~\cite{Guo:2012yt}. }
\label{fig.hist}
\end{center}
\end{figure}

To quantitatively estimate the uncertainties of $s_0$ and $\hat{m}_q$, we generate huge amounts of random configurations of the $R^{\rm (QCD)}$ by assigning ten-percent uncertainties to the OPE condensate parameters and the instanton size $\rho$. The value of $\kappa$ is allowed to randomly float between 2.0 and 4.0. In addition, we also generate large random samples of the spectral functions according to the results from Ref.~\cite{Guo:2012yt}, as explicitly shown in Fig.~\ref{exp-spectrum}. In Fig.~\ref{plot3d}, we show the typical dependence of the $\chi^2$ function with $s_0$ and $\hat{m}_q$ for a given set of $R^{\rm (QCD)}$ and $R^{\rm (phen)}$. There exists a minimum valley of the $\chi^2$ function in the $s_0$ and $\hat{m}_q$ plane, which implies the clear correlation of the two parameters in a single minimization procedure. Although the shape of the minimum valley sometimes looks moderately smooth, it is not difficult to find the minimum point in the valley, which is exactly the solutions of $s_0$ and $\hat{m}_q$ that we are looking for. Repeating the minimization procedures by using the random configurations of $R^{\rm (QCD)}$ and $R^{\rm (phen)}$ gives us a huge amount of solutions for $s_0$ and $\hat{m}_q$.

In  Fig.~\ref{fig.hist}, we show the histogram  distributions of the resulting parameters $s_0$ and $m_q$ (obtained from $\hat{m}_q$ via Eq.~\eqref{mq-mqhat}), which seem somewhat deviating the Gaussian-like distributions. After a scutinized investigation, we find that the non-Gaussian distributions in Fig.~\ref{fig.hist} are mainly caused by the $\rho$ parameter that describes the instanton size, although the Gaussian uncertainties are assumed for the $\rho$ during the error-bar analyses. However, by looking at the nontrivial depdence of the $\rho$ parameter in $R^{\rm (QCD)}$~\eqref{QCD-2}, it is not supprised to see the somewhat non-Gaussian outpus in Fig.~\ref{fig.hist}.
The resulting values of the $s_0$, $\hat{m}_q$ and $m_q$ with their statistical uncertainties contributed by the OPE parameters, instanton size, the $\kappa$ parameter and spectral functions from Ref.~\cite{Guo:2012yt} obtained in the Monte-Carlo-based sum rules are summarized in Table~\ref{MC-results}. 

\begin{table}
\caption{ Results from the Monte-Carlo based QCD sum rules: $s_0$, $\hat m_q$ and the scale-dependence light quark masses $m_q$. A conservative Borel window $0.37\,{\rm GeV}^{-2}<\tau<1.05\,{\rm GeV}^{-2}$ is taken here. See the text for details. The statistical uncertainties are contributed by the OPE condensates, the instanton size $\rho$, the parameter $\kappa$ and the spectral functions of Ref.~\cite{Guo:2012yt}, $i.e.$, the shaded areas shown in Fig.~\ref{exp-spectrum}. }
\begin{center}
\renewcommand\arraystretch{1.1}
\begin{tabular}{ c|c|c }
\hline\hline    $s_0/\text{GeV}^2$ & $\hat{m}_q /\text{MeV}$ & $m_q$ (2 GeV)/MeV\\
\hline  $3.10 \pm 0.11$  & $4.52 \pm 0.18$ & $3.44 \pm 0.14$ \\
\hline\hline
\end{tabular}
\end{center}
\label{MC-results}
\end{table}

To make a conservative estimation of the uncertainties, we also take  the spectral function with the parameters and amplitudes from Ref.~\cite{Guo:2011pa} to determine $s_0$ and the light-quark masses. The explicit results are found to be
\begin{eqnarray}\label{eq.mcs0mqprd11}
s_0= 2.94~{\rm GeV^2}\,, \quad \hat{m}_q= 4.10~{\rm MeV}\,, \quad m_q= 3.12~{\rm MeV}\,.
\end{eqnarray}
As in the SVZ sum-rule case, the deviations by using the different spectral functions calculated with the different parameters and amplitudes from Refs.~\cite{Guo:2011pa,Guo:2012yt} are considered as systematic uncertainties, which can be quadratically added to the statistical ones in Table~\ref{MC-results}. Therefore our final determination of the $s_0$ is
\begin{eqnarray}\label{eq.s0mcsr}
s_0= 3.10\pm 0.11 \pm 0.16~{\rm GeV^2}\,,
\end{eqnarray}
where the first error bar is the statistical one from Table~\ref{MC-results} and the second error bar, obtained by taking the difference of the central values in Table~\ref{MC-results} and Eq.~\eqref{eq.mcs0mqprd11}, stands for the systematical uncertainty caused by different spectral functions. Similar rules are also applied in the following discussions. The combined result for the RG-invariant light-quark mass $\hat{m}_q$ is
\begin{eqnarray}
\hat{m}_q= 4.52\pm 0.18 \pm 0.42 ~{\rm MeV}\,,
\end{eqnarray}
and the scale-dependent light-quark mass obtained at 2~GeV reads
\begin{eqnarray}\label{eq.mqmc}
m_q= 3.44\pm 0.14 \pm 0.32 ~{\rm MeV}\,,
\end{eqnarray}
which perfectly agrees with the determination from the SVZ sum rule in Eq.~\eqref{mq-SVZ-results} and is also nicely consistent with the PDG value~\cite{Zyla:2020zbs}
\begin{eqnarray}\label{PDG-value}
  m_q^{\text{PDG}}(2 \, \text{GeV}) = \frac{1}{2} (m_u + m_d) = 3.45 ^{+0.55}_{-0.15}\,\text{MeV}\,.
\end{eqnarray}

As shown in Fig.~\ref{ratio-tau}, the instanton term gives the largest nonperturbative contribution in the QCD spectral function. It is interesting to further study its roles in the determinations of the light-quark masses. If the contribution of the instanton is turned off,  we obtain in the Monte-Carlo QCD sum rule $s_0=2.28 \text{GeV}^{2}$, $m_q=3.77$ MeV for $\kappa = 3.0$, which can be compared with the values in Eqs.~\eqref{eq.s0mcsr} and \eqref{eq.mqmc} by taking all the contributions. In the SVZ QCD sum rule approach, we obtain similar values $s_0=2.25 \text{GeV}^{2}$, $m_q=3.78$ MeV for $\kappa = 3.0$ by simply turning off the instanton contributions. It is clear that the instanton tends to reduce the light-quark mass $m_q$ to the PDG average value. 

Comparing with the results from the Breit-Wigner-type spectral functions used in Ref.~\cite{Yuan:2017foa}, our predictions with more sophisticated scalar spectral functions and higher order QCD corrections reduce the magnitudes of $m_q$ and turns out to be perfectly  consistent with the PDG average value~\cite{Zyla:2020zbs}, which is mainly based on several lattice determinations.
It is worthy pointing out that our results in Eqs.~\eqref{mq-SVZ-results} and \eqref{eq.mqmc} obtained from the sum rules with the scalar currents are compatible with the recent determinations from the pseudoscalar channels~\cite{Dominguez:2018azt}.

\section{Conclusions}

In this work, we determine the average light-quark mass $m_q=(m_u+m_d)/2$ in the isoscalar scalar channel both from the SVZ and the Monte-Carlo-based QCD sum rules, where sophisticated phenomenological spectral functions from the chiral effective theory are used in our analysis. The phenomenological spectral functions, that are determined from the unitarized $U(3)$ chiral theory, coherently include the contributions from the $f_0(500), f_0(980)$ and $f_0(1370)$, together with the two-meson continuum effects. The present study gives an independent determination of the light-quark mass from the scalar QCD sum rules, compared to the lattice calculations and the sum rules in the pseudoscalar channels.

The prediction of the scale-dependent light-quark mass from the conventional SVZ sum-rule method is $m_q(2 \, \text{GeV})=(3.46^{+0.16}_{-0.22} \pm 0.33 )\, \text{MeV}$, where the first part of the uncertainties are caused by the OPE condensates, the instanton size, the $\kappa$ parameter, the Borel parameter $\tau$ and the threshold parameter $s_0$, and the second part of uncertainties are obtained by taking the two different sets of phenomenological spectral functions. Similarly our prediction from the Monte-Carlo-based QCD sum rule turns out to be $m_q (2\, \text{GeV}) = (3.44 \pm 0.14 \pm 0.32) \, \text{MeV}$, which is nicely consistent with the result from the SVZ approach and also remarkably agrees with the PDG value $3.45_{-0.15}^{+0.55}$~MeV. Our work provides a solid proof that a proper way to include $f_0(500)$ and $f_0(980)$ in the spectral functions is important to obtain sensible results in the QCD sum rules. The present theoretical formalism is also expected to be extended to the strange sector to determine the strange-quark mass.

\vspace{.7cm} {\bf Acknowledgments} \vspace{.3cm}

This work is supported in part by the National Natural Science Foundation of China(NSFC) under the Grant Nos.~11605039, 11975090, 11575052, Natural Science Foundation of Hebei Province with contract No. A2017205124, Science Foundation of Hebei Normal University under Contract No. L2016B08, and the Fundamental Research Funds for the Central Universities with contract No.~2242021R10099. Fang-Hui Yin and Wen-Ya Tian are equally contributed to this work.


\end{document}